\documentclass[journal]{IEEEtran} 		
\usepackage{ushort}
\usepackage[mathcal]{euscript}
\usepackage{amsmath}
\usepackage{amsthm}
\usepackage{amssymb}
\usepackage{graphicx}
\usepackage{epsfig}
\usepackage{booktabs}
\usepackage{colortbl}
\usepackage{amsmath}
\usepackage{xcolor}
\usepackage{graphicx}
\usepackage{algorithmic}
\usepackage{algorithm}
\usepackage{cite}
\usepackage{subfigure}

\providecommand{\aabs}[1]{\Big\lvert#1\Big\rvert}		
\providecommand{\mbf}[1]{\mathbf{#1}}						
\providecommand{\wt}[1]{\widetilde{#1}}					

\providecommand{\bsym}[1]{\boldsymbol{#1}}				
\providecommand{\bsymwt}[1]{\widetilde{\boldsymbol{#1}}}	
\providecommand{\mc}[1]{\mathcal{#1}}						
\providecommand{\mbb}[1]{\mathbb{#1}}	

\colorlet{tableheadcolor}{gray!25} 
\newcommand{\headcol}{\rowcolor{tableheadcolor}} %
\colorlet{tablerowcolor}{gray!10} 
\newcommand{\rowcol}{\rowcolor{tablerowcolor}} %
\newcommand{\topline}{\arrayrulecolor{black}\specialrule{0.1em}{\abovetopsep}{0pt}%
            \arrayrulecolor{tableheadcolor}\specialrule{\belowrulesep}{0pt}{0pt}%
            \arrayrulecolor{black}}
\newcommand{\midline}{\arrayrulecolor{tableheadcolor}\specialrule{\aboverulesep}{0pt}{0pt}%
            \arrayrulecolor{black}\specialrule{\lightrulewidth}{0pt}{0pt}%
            \arrayrulecolor{white}\specialrule{\belowrulesep}{0pt}{0pt}%
            \arrayrulecolor{black}}

\DeclareMathOperator*{\argmax}{arg \, max}
\DeclareMathOperator*{\argmin}{arg \, min}
\newtheorem{theorem}{Theorem}
\newtheorem{property}[theorem]{Property}

\IEEEoverridecommandlockouts

\begin{document}

\title{Target Detection Performance of Spectrum Sharing MIMO Radars}


\author{Awais~Khawar,~\IEEEmembership{}Ahmed~Abdelhadi,~\IEEEmembership{}and~T. Charles~Clancy~\IEEEmembership{}

\thanks{Awais Khawar (awais@vt.edu) is with Virginia Polytechnic Institute and State University, Arlington, VA, 22203.

Ahmed Abdelhadi (aabdelhadi@vt.edu) is with Virginia Polytechnic Institute and State University, Arlington, VA, 22203.

T. Charles Clancy (tcc@vt.edu) is with Virginia Polytechnic Institute and State University, Arlington, VA, 22203.

This work was supported by DARPA under the SSPARC program. Contract Award Number: HR0011-14-C-0027. The views expressed are those of the author and do not reflect the official
policy or position of the Department of Defense or the U.S. Government.

Distribution Statement A: Approved for public release; distribution is unlimited.

}

}

\maketitle

\begin{abstract}


Future wireless communication systems are envisioned to share radio frequency (RF) spectrum, with other services such as radars, in order to meet the growing spectrum demands. In this paper, we consider co-channel spectrum sharing between cellular systems and radars. We address the problem of target detection by radars that are subject to shape its waveform in a way that it doesn’t cause interference to cellular systems. We consider a multiple-input multiple-output (MIMO) radar and a MIMO cellular communication system with $\mc K$ base stations (BS). We propose a spectrum sharing algorithm which steers radar nulls, by projecting radar waveform onto the null space of interference channel, towards a `selected' BS, thus, protecting it from radar interference. This BS is selected, among $\mc K$ BSs, on the basis of guaranteeing minimum waveform degradation. We study target detection capabilities of this null-space projected (NSP) waveform and compare it with the orthogonal waveform. We derive the generalized 
likelihood ratio test (GLRT) for target detection and derive detector statistic for NSP and orthogonal waveform. The target detection performance for NSP and orthogonal waveform is studied theoretically and via Monte Carlo simulations.

%
%
%
%

\end{abstract}


\section{Introduction}\label{sec:intro}

Spectrum sharing between wireless communication systems and radars is an emerging area of research. In the past, spectrum has been shared primarily between wireless communication systems using opportunistic approaches by users equipped with cognitive radios \cite{Hay05}. This type of spectrum sharing has been made possible with the use of spectrum sensing \cite{YA09}, or geolocation databases \cite{MCM12}, or a combination of both in the form of radio environment maps (REM) \cite{ZMG07}. Some recent efforts have explored co-channel sharing approaches among secondary network entities, please see \cite{GPY14} and reference therein. However, in contrast, co-channel spectrum sharing between wireless systems and radars has received little attention thus far because of regulatory concerns. 

Spectrum policy regulators, in the past, have not allowed commercial wireless services in the radar bands, except in few cases, due to the fear of harmful interference from these services to radar systems. Recently, in the United States (U.S.), the Federal Communications Commission (FCC), has proposed to use the 3550-3650 MHz band for commercial broadband use \cite{FCC12_SmallCells}. The incumbents in this band are radar and satellite systems. The Commission has proposed that incumbents share this band with commercial communication systems. The Commission's spectrum sharing initiative is motivated by many factors including the President's National Broadband plan, which called to free up to 500 MHz of federal-held spectrum by 2020 \cite{FCC_NBP10}; surge in consumers' demand for access to mobile broadband, which operators can't meet with current spectrum allocation; the report on efficient spectrum utilization by President's Council of Advisers on Science and Technology (PCAST), which emphasized to share 1000 
MHz of government-held spectrum \cite{PCAST12}, and the low utilization of 3550-3650 MHz band by federal incumbents \cite{NTIA10}. 

In the future, when radio frequency (RF) spectrum will be shared among many different systems, e.g., radars and cellular systems, it is important to access the interference scenario. Of course, radars will cause interference to communication systems and vice versa if proper interference mitigation methods and novel sharing algorithms are not employed. In a study conducted by National Telecommunications and Information Administration (NTIA), it was observed that in order to protect commercial cellular communication systems, from high power radar signal, large exclusion zones are required \cite{NTIA10}. These exclusion zones cover a large portion of the U.S. where majority of the population lives and, thus, does not make a business case for commercial deployment in radar bands. In order to share radar bands for commercial operation we have to address the interference mitigation techniques at both the systems. In this work, we focus on interference caused by radar systems to communication system and propose 
methods to mitigate this interference.

The federal-commercial spectrum sharing is not a new practice. In fact, in the past, commercial wireless systems have shared government bands on a low transmit power basis, in order to protect incumbents from interference. An example of such a scenario is wireless local area network (WLAN) in the 5250-5350 MHz and 5470-5725 MHz radar bands \cite{FCC_5GHz_Radar06}. So the Commission's latest initiative to share 3.5 GHz radar band with small cells, i.e. wireless base stations operating on a low power, is in harmony with previous practices \cite{FCC12_SmallCells}. 

\subsection{Related Work}

On going research efforts have shown that there are numerous ways to share spectrum between radars and communication systems. 
Cooperative sensing based spectrum sharing approaches can be utilized where a radar's allocated bandwidth is shared with communication systems \cite{WMW+08, BNR12, SPC12}. A joint communication-radar platform can be envisioned in which a spectrally-agile radar performs an additional task of spectrum sensing and upon finding of unused frequencies it can change its operating frequency. In addition of spectrum sharing such a setting can enable co-located radar and communication system platforms for integrated communications and radar applications \cite{REM11, YZC12, SW11, FHH14}. Radar waveforms can be shaped in a way that they don't cause interference to communication systems \cite{CWX11, KAC+14DySPANProjection, KAC14DySPANWaveform,Mo_radar,Awais_MILCOM,SKA+14}. Moreover, database-aided sensing at communication systems \cite{PMM+14} and beamforming approaches at MIMO radars can be realized for spectrum sharing \cite{DH13}.


%
%
%
%
%
%
%
%
%
%
%
%
%
%
%

\subsection{Our Contributions}

%
%
%
%
%
%
%
%
%
%
%
%
%
%
%
%
%

The problem of target estimation, detection, and tracking lies at the heart of radar signal processing. This problem becomes critically important when we talk about sharing radar spectrum with other systems, say cellular systems. 
The regulatory work going on in the 3.5 GHz band to share radar spectrum with commercial systems is the motivation of this work. The focus of this work is to study target detection performance of a radar that is subject to share its spectrum with a cellular system. We consider spectrum sharing between a MIMO radar and a cellular system with many base stations. In our previous work, we have addressed the problem of radar waveform projection onto the null space of interference channel, in order to mitigate radar interference to communication system, and the problem of selection of interference channel for projection when we have a cellular system with $\mc K$ base stations \cite{KAC+14DySPANProjection}. In this work, we consider the same sharing scenario but study the target detection performance of radar for the null-space projected (NSP) waveform and compare it with that of the orthogonal radar waveform. We use the generalized likelihood ratio test (GLRT) for target detection and derive detector statistic 
for NSP and the orthogonal waveform.

%
%
%
%
%


\subsection{Notations}
Matrices are denoted by bold upper case letters, e.g. $\mbf A$, and vectors are denoted by bold lower case letters, e.g. $\mbf a$. Transpose, conjugate, and Hermitian operators are denoted by $(\cdot)^T, (\cdot)^*,$ and $(\cdot)^H$, respectively. Moreover, notations used throughout the paper are provided in Table \ref{tab:Notations} along with descriptions for quick reference. 

\renewcommand{\arraystretch}{1.5}
\begin{table}
\centering
\caption{Table of Notations}
\begin{tabular}{ll}
  \topline
  \headcol Notation &Description\\
  \midline
  		    $\mbf x(t)$					 &Transmitted radar (orthogonal) waveform\\
	\rowcol $\mbf a(\theta)$ 			 &Steering vector to steer signal to 											  target angle $\theta$\\
			$\mbf y(t)$					 &Received radar waveform\\
 	\rowcol	$\mbf R_{\mbf x}$ 			 &Correlation matrix of orthogonal 												  waveforms\\
			$\mbf s_j^{\text{UE}}(t)$ 	 &Signal transmitted by the 													$j^{\text{th}}$ UE in the 														$i^{\text{th}}$ cell\\
	\rowcol	$\mc L_i^{\text{UE}}$ 		&Total number of user equipments 												(UEs) in the $i^{\text{th}}$ cell \\
			$\mc K$						&Total number of BSs \\
	\rowcol $M$ 						&Radar transmit/receive antennas \\
			$N^{\text{BS}}$				&BS transmit/receive antennas  \\ 
	\rowcol	$\mbf H_i$					&$i^{\text{th}}$ interference channel\\
			$\mbf r_i(t)$				&Received signal at the 														$i^{\text{th}}$ BS\\
	\rowcol $\mbf P_i$					&Projection matrix for the 														$i^{\text{th}}$ channel \\
	
  \hline
\end{tabular}
\label{tab:Notations}
\end{table}

\subsection{Organization}
This paper is organized as
follows. Section~\ref{sec:model} discuss MIMO radar, target channel, orthogonal waveforms, interference channel, and our cellular system model. Moreover, it also discusses modeling and statistical assumptions. Section~\ref{sec:arch} discusses spectrum sharing between MIMO radar and cellular system and introduces sharing architecture and projection algorithms. Section \ref{sec:detection} presents the generalized likelihood ratio test (GLRT) for target detection and derives detector statistic for NSP and orthogonal waveform. Section~\ref{sec:sim}
discusses numerical results and compares performance of NSP and orthogonal waveform. Section~\ref{sec:conclusion} concludes the paper.

\section{System Model}\label{sec:model}

In this section, we introduce preliminaries of MIMO radar, point target in far-field, orthogonal waveforms, interference channel, and cellular system model. Moreover, we also discuss modeling and statistical assumptions along with RF environment assumptions used throughout the paper.

\subsection{Radar Model}\label{sec:radar}

The radar we consider in this paper is a colocated MIMO radar with $M$ transmit and receive antennas and is mounted on a ship. The colocated MIMO radar has antennas that have spacing on the order of half the wavelength. Another class of MIMO radar is widely-spaced MIMO radar where elements are widely-spaced which results in enhanced spatial diversity \cite{HBC08}. The colocated radar gives better spatial resolution and target parameter identification as compared to the widely-spaced radar \cite{LS07}.

\subsection{Target Model/Channel}
In this paper, we consider a point target model which is defined for targets having a scatterer with infinitesimal spatial extent. This model is a good assumption and is widely used in radar theory for the case when radar elements are colocated and their exists a large distance between the radar array and the target as compared to inter-element distance \cite{Sko08}. The signal reflected from a point target with unit radar cross-section (RCS) is mathematically represented by the Dirac delta function.

\subsection{Signal Model}

Let $\mbf x(t)$ be the signal transmitted from the $M$-element MIMO radar array, defined as
\begin{equation}
\mbf x(t)= \begin{bmatrix} x_1(t)e^{j \omega_c t} &x_2(t)e^{j \omega_c t} &\cdots &x_{M}(t)e^{j \omega_c t} \end{bmatrix}^T
\end{equation} 
where $x_k(t)e^{j \omega_c t}$ is the baseband signal from the $k^{\text{th}}$ transmit element, $\omega_c$ is the carrier angular frequency,
$t \in [0, T_o]$, with $T_o$ being the observation time. We define the transmit steering vector as
\begin{equation}
\mbf a_T(\theta) \triangleq \begin{bmatrix} e^{-j \omega_c \tau_{T_1}(\theta)} &e^{-j \omega_c \tau_{T_2}(\theta)} &\cdots &e^{-j \omega_c \tau_{T_{M}}(\theta)} \end{bmatrix}^T.
\label{eq:at}
\end{equation}
Then, the transmit-receive steering matrix can be written as
\begin{equation}
\mbf A (\theta) \triangleq \mbf a_R(\theta) \mbf a_T^T(\theta).
\end{equation}
Since, we are considering $M$ transmit and receive elements, we define $\mbf a(\theta) \triangleq \mbf a_T(\theta) \triangleq \mbf a_R(\theta)$. The signal received from a single point target, in far-field with constant radial velocity $v_r$, at an angle $\theta$ can be written as
\begin{equation}
\mbf y(t) = \alpha \, e^{-j \omega_D t}\mbf A(\theta) \,  \mbf x(t-\tau(t)) + \mbf n(t) \label{eqn:rxRadar}
\end{equation}
where $\tau(t)=\tau_{T_k}(t) + \tau_{R_l}(t)$, denoting the sum of propagation delays between the target and the $k^{\text{th}}$ transmit element and the $l^{\text{th}}$ receive element, respectively; $\omega_D$ is the Doppler frequency shift, and $\alpha$ represents the complex path loss including the propagation loss and the coefficient of reflection. 

\subsection{Modeling Assumptions}
In order to keep the analysis tractable we have made the following assumptions about our signal model: 
\begin{itemize}
\item The path loss $\alpha$ is assumed to be identical for all transmit and receive elements, due to the far-field assumption \cite{LS08}.

\item The angle $\theta$ is the azimuth angle of the target.

\item After compensating the range-Doppler parameters, we can simplify equation \eqref{eqn:rxRadar} as
\begin{equation}
\mbf y(t) = \alpha \, \mbf A(\theta) \,  \mbf x(t) + \mbf n(t). \label{eqn:rxRadarSimple}
\end{equation}
\end{itemize}

\subsection{Statistical Assumptions}
We make the following assumptions for our received signal model in equation \eqref{eqn:rxRadarSimple}:
\begin{itemize}
\item  $\theta$ and $\alpha$
are deterministic unknown parameters representing the target's direction of arrival and the complex amplitude of the target, respectively.

\item The noise vector $\mbf n(t)$ is independent, zero-mean complex Gaussian with known covariance matrix $\mbf R_{\mbf n} = \sigma^2_n \mbf I_M$, i.e. $\mbf n(t) \sim \mathcal{N}^c (\bsym 0_M,  \sigma^2_n \mbf I_M)$, where $\mathcal{N}^c$ denotes the complex Gaussian distribution.

\item With the above assumptions, the received signal model in equation \eqref{eqn:rxRadarSimple} has an independent complex Gaussian distribution, i.e., $\mbf y(t) \sim \mathcal{N}^c (\alpha \mbf A(\theta) \,  \mbf x(t),  \sigma^2_n \mbf I_M)$. 

\end{itemize}

\subsection{Orthogonal Waveforms}\label{sec:orthogonalWaveforms}
In this paper, we consider orthogonal waveforms transmitted by MIMO radars, i.e., 
\begin{equation}
\mbf R_{\mbf x} = \int_{T_o} \mbf x(t) \mbf x^H(t) dt  = \mbf I_M.
\end{equation}
The transmission of orthogonal signals gives MIMO radar advantages in terms of digital beamforming at the transmitter in addition to receiver, improved angular resolution, extended array aperture in the form of virtual arrays, increased number of resolvable targets, lower sidelobes, and lower probability of intercept as compared to coherent waveforms \cite{LS08}.

\subsection{Communication System}

In this paper, we consider a MIMO cellular system, with $\mc K$ base stations, each equipped with $N^{\text{BS}}$ transmit and receive antennas, with $i^{\text{th}}$ BS supporting $\mc L_i^{\text{UE}}$ user equipments (UEs). The UEs are also multi-antenna systems with  $N^{\text{UE}}$ transmit and receive antennas. If $\mbf s_j^{\text{UE}}(t)$ is the signals transmitted by the $j^{\text{th}}$ UE in the $i^{\text{th}}$ cell, then the received signal at the $i^{\text{th}}$ BS receiver can be written as
\begin{align}\label{eqn:RxPureComm}
\mbf r_i(t) &= \sum_j \mbf H_j^{N^{\text{BS}} \times N^{\text{UE}}} \mbf s_{j}^{\text{UE}}(t) + \mbf w(t) \\
&\quad \, \, \text{for} \, \, 1 \leq i \leq \mc K \, \, \text{and} \, \, 1 \leq j \leq \mc L_i^{\text{UE}} \nonumber
\end{align}
where 
$\mbf w(t)$ is the additive white Gaussian noise.


\subsection{Interference Channel}
In this section, we characterize the interference channel that exists between a MIMO cellular base station and a MIMO radar. In our paper, we are considering $\mc K$ cellular BSs that's why our model has $\mbf H_i, i=1, 2, \ldots, \mc K,$ interference channels, where the entries of $\mbf H_i$ are denoted by
\begin{equation}\label{Rx}
\mbf H_{i} = \begin{bmatrix} h_i^{(1,1)} &\cdots &h_i^{(1,M)}  
\\ \vdots  &\ddots &\vdots 
\\ h_i^{(N^{\text{BS}},1)}    &\cdots &h_i^{(N^{\text{BS}},M)} \end{bmatrix} \quad (N^{\text{BS}}  \times M)
\end{equation}
where $h_i^{(l,k)}$ denotes the channel coefficient from the $k^{\text{th}}$ antenna element at the MIMO radar to the $l^{\text{th}}$ antenna element at the $i^{\text{th}}$ BS. We assume that elements of $\mbf H_i$ are independent, identically distributed (i.i.d.) and circularly symmetric complex Gaussian random variables with zero-mean and unit-variance, thus, having a i.i.d. Rayleigh distribution. 

\subsection{Cooperative RF Environment}
In the wireless communications literature, it is usually assumed that the transmitter (mostly BSs) has channel state information either by feedback from the receiver (mostly UEs), in FDD systems \cite{TV05}, or transmitters can reciprocate the channel, in TDD systems \cite{TV05}. The feedback and reciprocity are valid and practical as long as the feedback has a reasonable overhead and coherence time of the RF channel is larger than the two-way communication time, respectively. 

In the case of radars sharing their spectrum with communications systems one way to get CSI from communication systems is through feedback. Since, a radar signal is treated as interference at a communication system, we can characterize the channel as interference channel and refer to information about it as interference-channel state information (ICSI). 

Spectrum sharing between radars and communications systems can be envisioned in two domains: military radars sharing spectrum with military communication systems, we call it \textit{Mil2Mil} sharing; another possibility is military radars sharing spectrum with commercial communication systems, we call it \textit{Mil2Com} sharing. In \textit{Mil2Mil} sharing, ICSI can be acquired by radars fairly easily as both systems belong to military. In \textit{Mil2Com} sharing, ICSI can be acquired by giving incentives to commercial communication system. The biggest incentive in this scenario is null-steering and protection from radar interference. Thus, regardless of the sharing scenario, \textit{Mil2Mil} or \textit{Mil2Com}, we have ICSI for the sake of mitigating radar interference at communication systems.

\section{Radar-Cellular System Spectrum Sharing}\label{sec:arch}

After introducing our radar and cellular system models we can now discuss the spectrum sharing scenario between radar and cellular system. In our sharing architecture MIMO radar and cellular systems are the co-primary users of the 3550-3650 MHz band under consideration. In the following sections, we will discuss the architecture of spectrum sharing problem which is followed by our spectrum sharing algorithm.

\begin{figure}
\centering
	\includegraphics[width=3.4in]{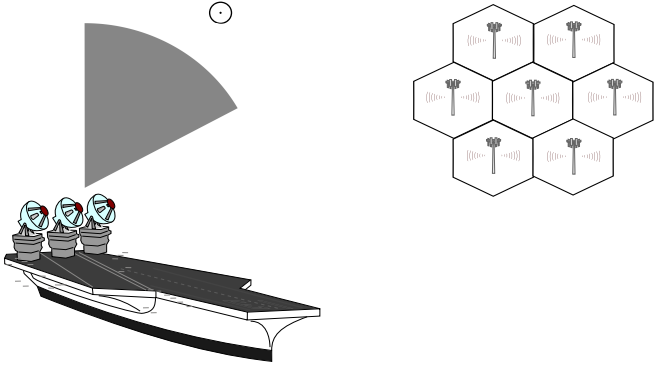} 
	\caption{Spectrum Sharing Scenario: A seaborne MIMO radar detecting a point target while simultaneously sharing spectrum with a MIMO cellular system without causing interference to the cellular system.}
		\label{fig:scenario}
\end{figure}

\subsection{Architecture}

We illustrate our coexistence scenario in Figure \ref{fig:scenario} where the maritime MIMO radar is sharing $\mc K$ interference channels with the cellular system. Considering this scenario, the received signal at the $i^{\text{th}}$ BS receiver can be written as
\begin{align}\label{eqn:RxRadarComm}
\mbf r_i(t) = \mbf H_i^{N^{\text{BS}} \times M} \mbf x(t) + \sum_j \mbf H_j^{N^{\text{BS}} \times N^{\text{UE}}} \mbf s_{j}^{\text{UE}}(t) + \mbf w(t).
\end{align}
The goal of the MIMO radar is to map $\mbf x(t)$ onto the null-space of $\mbf H_i$ in order to avoid interference to the $i^{\text{th}}$ BS, i.e., $\mbf H_i \mbf x(t) = \bsym 0$, so that $\mbf r_i(t)$ has equation \eqref{eqn:RxPureComm} instead of equation \eqref{eqn:RxRadarComm}. 


\subsection{Projection Matrix}

In this section, we define the projection algorithm which projects radar signal onto the null space of interference channel $\mbf H_i$. Assuming, the MIMO radar has channel state information of all $\mbf H_i$ interference channels, through feedback, in \textit{Mil2Mil} or \textit{Mil2Com} scenario, we can perform singular value decomposition (SVD) to find the null space and then construct a projector matrix. We proceed by first finding SVD of $\mbf H_i$, i.e., 
\begin{equation}
\mbf H_i = \mbf U_i \bsym \Sigma_i \mbf V_i^H.
\end{equation}
Now, let us define 
\begin{equation}
\bsymwt \Sigma_i \triangleq \text{diag} (\wt \sigma_{i,1}, \wt \sigma_{i,2}, \ldots, \wt \sigma_{i,p})
\end{equation}
where $p \triangleq \min (N^{\text{BS}},M)$ and 
$\wt \sigma_{i,1} > \wt \sigma_{i,2} > \cdots > \wt \sigma_{i,q} > \wt \sigma_{i,q+1} = \wt \sigma_{i,q+2} = \cdots = \wt \sigma_{i,p} = 0$ are the singular values of $\mbf H_i$. Next, we define
\begin{equation}
\bsymwt {\Sigma}_i^\prime \triangleq \text{diag} (\wt \sigma_{i,1}^\prime,\wt \sigma_{i,2}^\prime, \ldots, \wt \sigma_{i,M}^\prime)
\end{equation}
where
\begin{align}
\wt \sigma_{i,u}^\prime \triangleq
\begin{cases}
0, \quad \text{for} \; u \leq q,\\
1, \quad \text{for} \; u > q.
\end{cases}
\end{align}
Using above definitions we can now define our projection matrix, i.e.,
\begin{equation}\label{eqn:ProjDefinition}
\mbf P_i \triangleq \mbf V_i \bsymwt \Sigma_i^\prime \mbf V_i^H.
\end{equation}
In order to show that $\mbf P_i$ is a valid projection matrix we prove two results on projection matrices below.
\begin{property}
$\mbf P_i \in \mbb C^{M \times M}$ is a projection matrix if and only if $\mbf P_i = \mbf P_i^H = \mbf P_i^2$.
\end{property}
\begin{proof}
Let's start by showing the `only if' part. First, we show $\mbf P_i = \mbf P_i^H$. Taking Harmition of equation \eqref{eqn:ProjDefinition} we have
\begin{equation}\label{eqn:Projhermitian}
\mbf P_i^H = (\mbf V_i \bsymwt \Sigma_i^\prime \mbf V_i^H)^H = \mbf P_i.
\end{equation}
Now, squaring equation \eqref{eqn:ProjDefinition} we have
\begin{equation}\label{eqn:ProjSquare}
\mbf P_i^2 = \mbf V_i \bsymwt \Sigma_i \mbf V^H_i \times \mbf V_i \bsymwt \Sigma_i \mbf V^H_i = \mbf P_i
\end{equation}
where above equation follows from $ \mbf V^H_i \mbf V_i = \mbf I$ (since they are orthonormal matrices) and $(\bsymwt \Sigma_i^\prime)^2 =\bsymwt \Sigma_i^\prime$ (by construction). From equations \eqref{eqn:Projhermitian} and \eqref{eqn:ProjSquare} it follows that $\mbf P_i = \mbf P_i^H = \mbf P_i^2$.
Next, we show $\mbf P_i$ is a projector by showing that if $\mbf v \in$ range ($\mbf P_i$), then $\mbf P_i \mbf v = \mbf v$, i.e., for some $\mbf w, \mbf v = \mbf P_i \mbf w$, then
\begin{equation}
\mbf P_i \mbf v = \mbf P_i (\mbf P_i \mbf w) = \mbf P_i^2 \mbf w = \mbf P_i \mbf w = \mbf v.
\end{equation}
Moreover, $\mbf P_i \mbf v - \mbf v \in$ null($\mbf P_i$), i.e., 
\begin{equation}
\mbf P_i (\mbf P_i \mbf v - \mbf v) = \mbf P_i^2 \mbf v -\mbf P_i \mbf v = \mbf P_i \mbf v - \mbf P_i \mbf v = \mbf 0.
\end{equation}
This concludes our proof.

\end{proof}

\begin{property}
$\mbf P_i \in \mbb C^{M\times M}$ is an orthogonal projection matrix onto the null space of  $\mbf H_i \in \mbb C^{N^{\text{BS}}\times M}$.
\end{property}
\begin{proof}
Since $\mbf P_i = \mbf P_i^H$, we can write 
\begin{equation}
\mbf H_i \mbf P_i^H = \mbf U_i \bsymwt \Sigma_i \mbf V_i^H \times \mbf V_i \bsymwt \Sigma_i^\prime \mbf V^H_i = \bsym 0.
\end{equation}
The above results follows from noting that $\bsymwt \Sigma_i \bsymwt \Sigma_i^\prime = \bsym 0$ by construction.
\end{proof}

In this paper, we are dealing with $\mc K$ interference channels. Therefore, we need to select the interference channel which results in least degradation of radar waveform in a minimum norm sense, i.e.,
\begin{align}
i_{\text{min}} &\triangleq \argmin_{1 \leq i \leq \mc K} \Big| \Big|\mbf P_i \mbf x(t) - \mbf x(t)\Big|\Big|_2 \\
\breve{\mbf P} &\triangleq \mbf P_{i_{\text{min}}}.
\end{align}
Once we have selected our projection matrix it is straight forward to project radar signal onto the null space of interference channel via
\begin{equation}
\breve{\mbf x}(t) = \breve{\mbf P} \: \mbf x(t).
\end{equation}
The correlation matrix of our NSP waveform is given as
\begin{equation}
\mbf R_{\breve{\mbf x}} = \int_{T_o} \breve{\mbf x}(t) \breve{\mbf x}^H(t) dt
\end{equation}
which is no longer identity and its rank depends upon the rank of the projection matrix.

\subsection{Spectrum Sharing and Projection Algorithms}

The process of spectrum sharing by forming projection matrices and selecting interference channels is executed with the help of Algorithms \eqref{alg:M2C_Radar} and \eqref{alg:Proj}. First, at each pulse repetition interval (PRI), the radar obtains ICSI of all $\mc K$ interference channels. This information is sent to Algorithm \eqref{alg:Proj} for the calculation of null spaces and formation of projection matrices. Algorithm \eqref{alg:M2C_Radar} process $\mc K$ projection matrices, received from Algorithm \eqref{alg:Proj}, to find the projection matrix which results in least degradation of radar waveform in a minimum norm sense. This step is followed by the projection of radar waveform onto the null space of the selected BS, i.e, the BS to the corresponding selected projection matrix, and waveform transmission.

\begin{algorithm}
\caption{Spectrum Sharing Algorithm}\label{alg:M2C_Radar}
\begin{algorithmic}
\LOOP
	\FOR{$i=1:\mc K$}
		\STATE{Get CSI of $\mbf H_i$ through feedback from the $i^{\text{th}}$ BS.}
		\STATE{Send $\mbf H_i$ to Algorithm \eqref{alg:Proj} for the formation of projection matrix $\mbf P_i$.}
		\STATE{Receive the $i^{\text{th}}$ projection matrix $\mbf P_i$ from Algorithm \eqref{alg:Proj}.}
	\ENDFOR
	\STATE{Find {$i_{\text{min}} = \argmin\limits_{1 \leq i \leq \mc K} \Big| \Big|\mbf P_i \mbf x(t) - \mbf x(t)\Big|\Big|_2$.}}
\STATE{Set $\breve{\mbf P} = \mbf P_{i_{\text{min}}}$ as the desired projector.}
\STATE{Perform null space projection, i.e., $\breve{\mbf x}(t) = \breve{\mbf P} \mbf x(t)$.}
\ENDLOOP
\end{algorithmic}
\end{algorithm}

\begin{algorithm}
\caption{Projection Algorithm}\label{alg:Proj}
\begin{algorithmic}
\IF {$\mbf H_i$ received from Algorithm \eqref{alg:M2C_Radar}}
	\STATE{Perform SVD on $\mbf H_i$ (i.e. $\mbf H_i=\mbf U_i \bsym \Sigma_i \mbf V_i^H$)}
	\STATE{Construct $\bsymwt \Sigma_i = \text{diag} (\wt \sigma_{i,1}, \wt \sigma_{i,2}, \ldots, \wt \sigma_{i,p})$}
	\STATE{Construct $\bsymwt {\Sigma}_i^\prime = \text{diag} (\wt \sigma_{i,1}^\prime,\wt \sigma_{i,2}^\prime, \ldots, \wt \sigma_{i,M}^\prime)$}
	\STATE{Setup projection matrix $\mbf P_i = \mbf V_i \bsymwt \Sigma_i^\prime \mbf V_i^H$.}	
\STATE{Send $\mbf P_i$ to Algorithm \eqref{alg:M2C_Radar}.}	
\ENDIF 

\end{algorithmic}
\end{algorithm}

\section{Statistical Decision Test for Target Detection}\label{sec:detection}

\begin{figure}
\centering
	\includegraphics[width=3.4in]{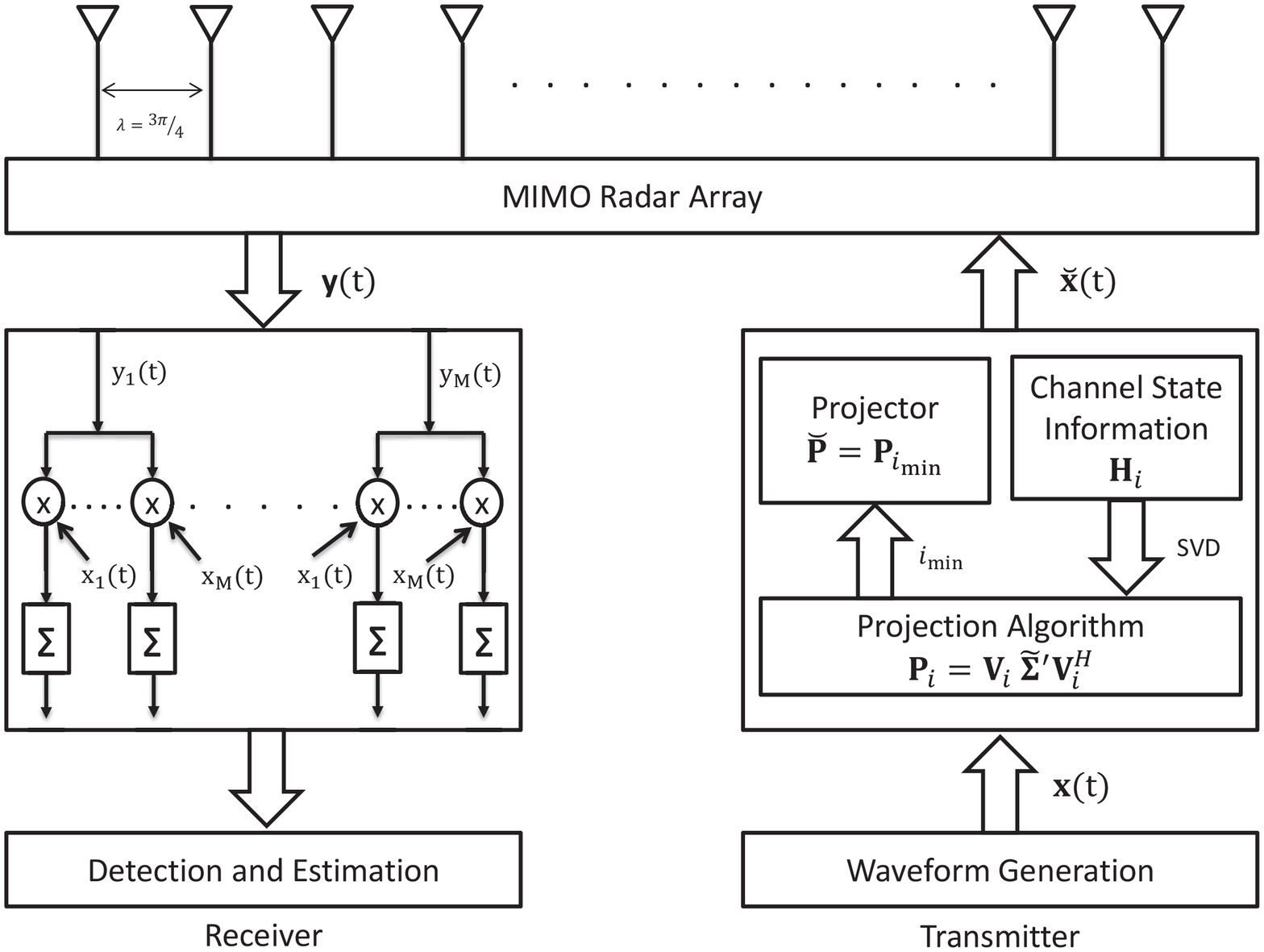} 
	\caption{Block diagram of spectrum sharing radar. The transmitter is modified to perform the functions of ICSI collection, projection matrix formation, interference channel selection, and radar waveform projection on to the selected interference channel for spectrum sharing. On the other hand, the receiver is a traditional radar receiver performing functions of parameter detection and estimation on radar returns.}
		\label{fig:block}
\end{figure}

In this section, we develop a statistical decision test for target illuminated with the orthogonal radar waveforms and the NSP projected radar waveforms. The goal is to compare the performance of the two waveforms by looking at the test decision on whether the target is present or not in the range-Doppler cell of interest. We present a system-level architecture of the spectrum sharing radar in Figure \ref{fig:block}. In our architecture, the transmitter performs the functions of waveform generation, channel selection, and projection; and the receiver performs the functions of signal detection and estimation.



For target detection and estimation, we proceed by constructing a hypothesis test where we seek to choose between two hypothesis: the null hypothesis $\mc H_0$ which represents the case when the target is absent or the alternate hypothesis $\mc H_1$ which represents the case when the target is present.  The hypothesis for a single target model in equation \eqref{eqn:rxRadarSimple} can be written as
\begin{align}\label{eqn:Hypo}
\mbf y(t) = 
\begin{cases}
\mc H_{1}:\alpha \, \mbf A(\theta) \,  \mbf x(t) + \mbf n(t), & 0 \leq t \leq T_o, \\
\mc H_{0}:\mbf n(t), & 0 \leq t \leq T_o. 
\end{cases}
\end{align}
Since, $\theta$ and $\alpha$ are unknown, but deterministic, we use the generalize likelihood ratio test (GLRT). The advantage of using GLRT is that we can replace the unknown parameters with their maximum likelihood (ML) estimates. The ML estimate of $\alpha$ and $\theta$ are found for various signal models, targets, and interference sources in \cite{LS08,BT06} when using orthogonal signals. In this paper, we consider a simpler model with one target and no interference sources in order to study the impact of NSP on target detection in a tractable manner. Therefore, we present a simpler derivation of ML estimation and GLRT.

The received signal model in equation \eqref{eqn:rxRadarSimple} can be written as
\begin{equation}\label{eqn:rxRadarKL}
\mbf y(t) = \mbf Q(t, \theta) \alpha + \mbf n(t)
\end{equation}
where
\begin{equation}
\mbf Q(t, \theta) = \mbf A(\theta)  \mbf x (t) \label{eqn:Q_def}.
\end{equation}

We use Karhunen-Lo\`eve expansion for derivation of the log-likelihood function for estimating $\theta$ and $\alpha$. Let $\Omega$ denote the space of the elements of $\{\mbf y(t)\}, \{\mbf Q(t,\theta) \},$ and $\{\mbf n(t)\}$. Moreover, let $\psi_z$, $z=1,2,\ldots,$ be an orthonormal basis function of $\Omega$ satisfying 
\begin{equation}
< \psi_z(t),\psi_{z'}(t)> = \int_{T_0} \psi_z(t),\psi_{z'}^*(t) =\delta_{zz'}
\end{equation}
where $\delta_{zz'}$ is the Kr\"onecker delta function. Then, the following series can be used to expand the processes, $\{\mbf y(t)\}, \{\mbf Q(t,\theta) \},$ and $\{\mbf n(t)\}$, as
\begin{align}
\mbf y(t) &= \sum_{z=1}^\infty \mbf y_z \psi_z(t) \\
\mbf Q(t,\theta) &= \sum_{z=1}^\infty \mbf Q_z(\theta) \psi_z(t) \\
\mbf n(t) &= \sum_{z=1}^\infty \mbf n_z \psi_z(t)
\end{align}
where $\mbf y_z, \mbf Q_z$, and $\mbf n_z$ are the coefficients in the  Karhunen-Lo\`eve expansion of the considered processes obtained by taking the corresponding inner product with basis function $\phi_z(t)$. Thus, 
an equivalent discrete model of equation \eqref{eqn:rxRadarKL} can be obtained as
\begin{equation}
\mbf y_z = \mbf Q_z(\theta) \alpha + \mbf n_z, \quad z=1,2,\ldots
\end{equation}
For white circular complex Gaussian processes, i.e, $\mbb E [ \mbf n(t) \mbf n(t-\tau(t))] = \sigma^2_n \mbf I_M \delta(\tau(t))$, the sequence $\{n_z\}$ is i.i.d. and $\mbf n_z \sim \mathcal N^c (\bsym 0_M, \sigma^2_n \mbf I_M)$. Thus, we can express the log-likelihood function as
\begin{equation}
L_{\mbf y} (\theta, \alpha) = \sum_{z=1}^\infty \Bigg(-M \log (\pi \sigma^2_n ) - \frac{1}{\sigma^2_n} \aabs{\aabs{\mbf y_z - \mbf Q_z(\theta) \alpha}}^2 \Bigg).
\end{equation}
Maximizing with respect to $\alpha$ yields
\begin{equation}
L_{\mbf y} (\theta,\hat \alpha) = \Gamma -  \frac{1}{\sigma^2_n} \Big( E_{\mbf y \mbf y} - \mbf e^H_{\mbf Q \mbf y} \mbf E_{\mbf Q \mbf Q}^{-1} \mbf e_{\mbf Q \mbf y}  \Big) \label{eqn:LyMAx}
\end{equation}
where
\begin{align}
\Gamma &\triangleq -M \log (\pi \sigma^2_n )\\
E_{\mbf y \mbf y} &\triangleq \sum_{z=1}^\infty \aabs{\aabs{\mbf y_z}}^2 \label{eqn:KL_Eyy}\\
\mbf e_{\mbf Q \mbf y} &\triangleq \sum_{z=1}^\infty \mbf Q_z^H \mbf y_z \label{eqn:KL_eHy} \\
\mbf E_{\mbf Q \mbf Q}^{-1} &\triangleq \sum_{z=1}^\infty \mbf Q_z^H \mbf Q_z. \label{eqn:KL_EHH}
\end{align}
Note that, in equation \eqref{eqn:LyMAx}, apart from the constant $\Gamma$, the remaining summation goes to infinity. However, due to the non-contribution of higher order terms in the estimation of $\theta$ and $\alpha$ the summation can be finite. Using the identity
\begin{equation}
\int_{T_o} \mbf v_1(t) \mbf v_2^H(t) dt = \sum_{z=1}^\infty \mbf v_{1z} \mbf v_{2z}^H
\end{equation}
for $\mbf v_i(t) = \sum_{z=1}^\infty \mbf v_{1z} \psi_z(t), i=1,2,$ equations  \eqref{eqn:KL_Eyy}-\eqref{eqn:KL_EHH} can be written as
\begin{align}
E_{\mbf y \mbf y} &\triangleq \int_{T_o}  \aabs{\aabs{\mbf y(t)}}^2 dt\\
\mbf e_{\mbf Q \mbf y} &\triangleq \int_{T_o}  \mbf Q^H(t, \theta) \mbf y(t) dt \\
\mbf E_{\mbf Q \mbf Q} &\triangleq \int_{T_o} \mbf Q^H(t,\theta) \mbf Q(t, \theta) dt.
\end{align}
Using the definition of $\mbf Q(t,\theta)$ in equation \eqref{eqn:Q_def}, we can write the $f^{\text{th}}$ element of $\mbf e_{\mbf Q \mbf y}$ as
\begin{equation}\label{eqn:eQy}
[\mbf e_{\mbf Q \mbf y}]_f = \mbf a^H(\theta_f) \mbf E^T \mbf a(\theta_f)
\end{equation}
where
\begin{equation}
\mbf E = \int_{T_o} \mbf y(t) \mbf x^H(t) dt.
\end{equation}
Similarly, we can write the ${fg}^{\text{th}}$ element of $\mbf E_{\mbf Q \mbf Q}$ as
\begin{equation}\label{eqn:EQQ}
[\mbf E_{\mbf Q \mbf Q}]_{fg} = \mbf a^H (\theta_f)\mbf a(\theta_g) \mbf a^H(\theta_f) \mbf R_x^T \mbf a(\theta_g).
\end{equation}
Since, $\mbf e_{\mbf Q \mbf y}$ and $\mbf E_{\mbf Q \mbf Q}$ are independent of the received signal, the sufficient statistic to calculate $\theta$ and $\alpha$ is given by $\mbf E$. Using equation \eqref{eqn:eQy}-\eqref{eqn:EQQ} we can write the ML estimate in matrix-vector form as
\begin{equation}
L_{\mbf y}(\hat{\theta}_{\text{ML}}) = \argmax\limits_{\theta} \frac{\left| \mathbf{a}^H(\hat{\theta}_{\text{ML}}) \mbf E \mathbf{a}^*(\hat{\theta}_{\text{ML}})\right|^2}{M \mathbf{a}^H(\hat{\theta}_{\text{ML}}) \mathbf{R}_{\mbf x}^T \mathbf{a}(\hat{\theta}_{\text{ML}})}\cdot
\end{equation}

Then, the GLRT for our hypothesis testing model in equation \eqref{eqn:Hypo}  is given as
\begin{equation}
L_{\mbf y} = \max_{\theta, \alpha} \{ \log f_{\mbf y}(\mbf y,\theta,\alpha;\mc H_1)\} - \log f(\mbf y;\mc H_0) \mathop{\gtrless}^{\mc H_1}_{\mc H_0} \delta
\end{equation}
where $ f_{\mbf y}(\mbf y,\theta,\alpha;\mc H_1)$ and $ f(\mbf y;\mc H_0)$ are the probability density functions of the received signal under hypothesis $\mc H_1$ and $\mc H_0$, respectively. Hence, the GLRT can be expressed as
\begin{equation}
\label{eq:ML}
L_{\mbf y}(\hat{\theta}_{\text{ML}}) = \argmax\limits_{\theta} \frac{\left| \mathbf{a}^H(\hat{\theta}_{\text{ML}}) \mbf E \mathbf{a}^*(\hat{\theta}_{\text{ML}})\right|^2}{M \mathbf{a}^H(\hat{\theta}_{\text{ML}}) \mathbf{R}_{\mbf x}^T \mathbf{a}(\hat{\theta}_{\text{ML}})} \mathop{\gtrless}^{\mc H_1}_{\mc H_0} \delta.
\end{equation}

The asymptotic statistics of $L(\hat{\theta}_{\text{ML}})$ for both the hypothesis is given by \cite{Kay98}
\begin{align}
L(\hat{\theta}_{\text{ML}}) \sim 
\begin{cases}
\mc H_{1}: \chi^2_2(\rho), \\
\mc H_{0}:\chi^2_2,
\end{cases}
\end{align}
where 
\begin{itemize}
\item $\chi^2_2(\rho)$ is the noncentral chi-squared distributions with two degrees of freedom,
\item $\chi^2_2$ is the central chi-squared distributions with two degrees of freedom,

\item and $\rho$ is the noncentrality parameter, which is given by
\begin{equation}
\rho = \dfrac{|\alpha|^2}{\sigma^2_n} | \mathbf{a}^H({\theta}) \mathbf{R}_{\mbf x}^T \mathbf{a}({\theta})|^2.
\end{equation}
\end{itemize}
For the general signal model, we set $\delta$ according to a desired probability of false alarm $P_{\text{FA}}$, i.e.,
\begin{align}
P_{\text{FA}} &= P(L(\mbf y) > \delta | \mc H_0) \\
\delta &= \mathcal{F}^{-1}_{\chi^2_2} (1-P_{\text{FA}})
\end{align}
where $ \mathcal{F}^{-1}_{\chi^2_2}$ is the inverse central chi-squared distribution function with two degrees of freedom. The probability of detection is given by
\begin{align}
P_{\text{D}} &= P(L(\mbf y) > \delta | \mc H_1) \\
P_{\text{D}} &=1 - \mathcal{F}_{\chi^2_2(\rho)} \left( \mathcal{F}^{-1}_{\chi^2_2} (1-P_{\text{FA}}) \right)
\end{align}
where $\mathcal{F}_{\chi^2_2(\rho)}$ is the noncentral chi-squared distribution function with two degrees of freedom and noncentrality parameter $\rho$.

\subsection{$P_D$ for Orthogonal Waveforms}
For orthogonal waveforms $\mathbf{R}_{\mbf x}^T = \mbf I_M$, therefore, the GLRT can be expressed as
\begin{equation}
\label{eqn:ML_Orthog}
L_{\text{Orthog}}(\hat{\theta}_{\text{ML}}) = \frac{\left| \mathbf{a}^H(\hat{\theta}_{\text{ML}}) \mbf E \mathbf{a}^*(\hat{\theta}_{\text{ML}})\right|^2}{M \mathbf{a}^H(\hat{\theta}_{\text{ML}}) \mathbf{a}(\hat{\theta}_{\text{ML}})} \mathop{\gtrless}^{\mc H_1}_{\mc H_0} \delta_{\text{Orthog}}
\end{equation}
and the statistics of $L(\hat{\theta}_{\text{ML}})$ for this case is
\begin{align}
L_{\text{Orthog}}(\hat{\theta}_{\text{ML}}) \sim 
\begin{cases}
\mc H_{1}: \chi^2_2(\rho_{\text{Orthog}}), \\
\mc H_{0}:\chi^2_2,
\end{cases}
\end{align}
where 
\begin{equation}
\rho_{\text{Orthog}} = \dfrac{M^2 |\alpha|^2}{\sigma^2_n} \cdot 
\end{equation}
We set $\delta_{\text{Orthog}}$ according to a desired probability of false alarm $P_{\text{PF-Orthog}}$, i.e.,
\begin{equation}
\delta_{\text{Orthog}} = \mathcal{F}^{-1}_{\chi^2_2} (1-P_{\text{PF-Orthog}})
\end{equation}
and then the probability of detection for orthogonal waveforms is given by
\begin{equation}
P_{\text{D-Orthog}} =1 - \mathcal{F}_{\chi^2_2(\rho_{\text{Orthog}})} \left( \mathcal{F}^{-1}_{\chi^2_2} (1-P_{\text{PF-Orthog}}) \right).
\end{equation}

\subsection{$P_D$ for NSP Waveforms}
For spectrum sharing waveforms $\mbf{R}_{{\mbf x}}^T = \mbf{R}_{\breve{\mbf x}}^T$, therefore, the GLRT can be expressed as
\begin{equation}
\label{eqn:ML_NSP}
L_{\text{NSP}}(\hat{\theta}_{\text{ML}}) = \frac{\left| \mathbf{a}^H(\hat{\theta}_{\text{ML}}) \mbf E \mathbf{a}^*(\hat{\theta}_{\text{ML}})\right|^2}{M \mathbf{a}^H(\hat{\theta}_{\text{ML}})\mathbf{R}_{\breve{\mbf x}}^T \mathbf{a}(\hat{\theta}_{\text{ML}})} \mathop{\gtrless}^{\mc H_1}_{\mc H_0} \delta_{\text{NSP}}
\end{equation}
and the statistics of $L(\hat{\theta}_{\text{ML}})$ for this case is
\begin{align}
L_{\text{NSP}}(\hat{\theta}_{\text{ML}}) \sim 
\begin{cases}
\mc H_{1}: \chi^2_2(\rho_{\text{NSP}}), \\
\mc H_{0}:\chi^2_2,
\end{cases}
\end{align}
where 
\begin{equation}
\rho_{\text{NSP}}=\dfrac{|\alpha|^2}{\sigma^2_n} | \mathbf{a}^H({\theta}) \mathbf{R}_{\breve{\mbf x}}^T \mathbf{a}({\theta})|^2.
\end{equation}
We set $\delta_{\text{NSP}}$ according to a desired probability of false alarm $P_{\text{PF-NSP}}$, i.e.,
\begin{equation}
\delta_{\text{NSP}} = \mathcal{F}^{-1}_{\chi^2_2} (1-P_{\text{PF-NSP}})
\end{equation}
and then the probability of detection for orthogonal waveforms is given by
\begin{equation}
P_{\text{D-NSP}} =1 - \mathcal{F}_{\chi^2_2(\rho_{\text{NSP}})} \left( \mathcal{F}^{-1}_{\chi^2_2} (1-P_{\text{PF-NSP}}) \right).
\end{equation}

\section{Numerical Results}\label{sec:sim}

In order to study the detection performance of spectrum sharing MIMO radars, we carry out Monte Carlo simulation using the radar parameters mentioned in Table \ref{tab1}. At each run of Monte Carlo simulation we generate $\mc K$ Rayleigh interference channels each with dimensions $N^{\text{BS}} \times M$, calculate their null spaces and construct corresponding projection matrices using Algorithm \eqref{alg:Proj}, determine the best channel to perform projection of radar signal using Algorithm \eqref{alg:M2C_Radar}, transmit NSP signal, estimate parameters $\theta$ and $\alpha$ from the received signal, and calculate the probability of detection
for orthogonal and NSP waveforms.

\renewcommand{\arraystretch}{1.5}
\begin{table}[h]
\centering
\caption{MIMO Radar System Parameters}
\begin{tabular}{lll}
  \topline
  \headcol Parameters &Notations &Values\\
  \midline
  		    Radar/Communication System RF band &- &$3550-3650$ MHz \\
	\rowcol Radar antennas & $M$ &8, 4 \\
			Communication System Antennas & $N^{\text{BS}}$ &2 \\ 
 	\rowcol	Carrier frequency &$f_c$ &3.55 GHz \\
			Wavelength &$\lambda$ &8.5 cm \\
	\rowcol	Inter-element antenna spacing &$3\lambda/4$ &6.42 cm \\
			Radial velocity & $v_r$ &2000 m/s \\
	\rowcol	Speed of light &$c$ & 3 $\times$ $10^8$ m/s\\
			Target distance from the radar & $r_0$ &500 Km \\
	\rowcol Target angle &$\theta$ & $\hat \theta$ \\
			Doppler angular frequency &$\omega_D$  & $2 \omega_c v_r/c $  \\
	\rowcol	Two way propagation delay &$\tau_r$ & $2 r_0/c$ \\
			Path loss &$\alpha$  & $\hat \alpha$ \\
	
  \hline
\end{tabular}
\label{tab1}
\end{table}

\subsection{Performance of Algorithms \eqref{alg:M2C_Radar} and \eqref{alg:Proj}}
\begin{figure}
\centering
\subfigure[Probability of detection when $N^{\text{BS}}<M$. As an example we use $2 \times 4$ configuration. Note that 6 dB to 13 dB of additional gain in SNR is required to detect target with 90\% probability, depending upon the NSP waveform transmitted.]{
\includegraphics[width=\linewidth]{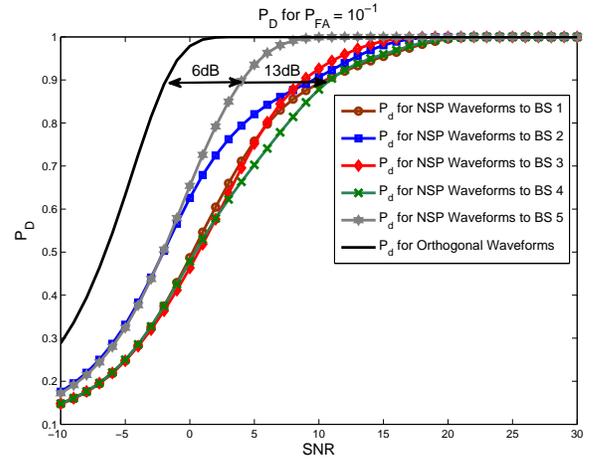}
\label{fig:PdH4x2_BS_5}
}
\subfigure[Probability of detection when $N^{\text{BS}}\ll M$. As an example we use $2 \times 8$ configuration. Note that 3 dB to 5 dB of additional gain in SNR is required to detect target with 90\% probability, depending upon the NSP waveform transmitted.]{
\includegraphics[width=\linewidth]{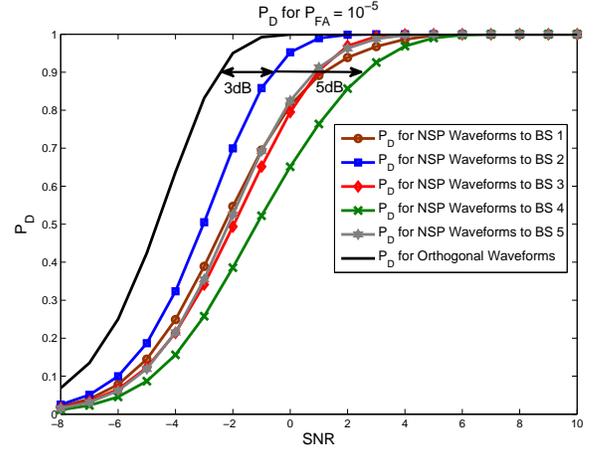}
\label{fig:PdH8x2_BS_5}
} 
\caption{Performance of Algorithms \eqref{alg:M2C_Radar} and \eqref{alg:Proj}: Using our spectrum sharing and projection algorithms, we can select interference channel for radar signal projection to maximize detection probability and minimize gain in SNR required as a result of NSP of radar waveforms. For example, Algorithms \eqref{alg:M2C_Radar} and \eqref{alg:Proj} select BS\#5 and BS\#2 for $N^{\text{BS}}<M$ and $N^{\text{BS}}<M$ cases, respectively, as they require minimum additional gain in SNR.}
\label{fig:MultiBS}
\end{figure}

In Figure \ref{fig:MultiBS}, we demonstrate the use of Algorithms \eqref{alg:M2C_Radar} and \eqref{alg:Proj} in improving target detection performance when multiple BSs are present in detection space of radar and the radar has to reliably detect target while not interfering with communication system of interest. As an example, we consider a scenario with five BSs and the radar has to select a projection channel which minimizes degradation in its waveform, thus, maximizing its probability of detection of the target.

In Figure \ref{fig:PdH4x2_BS_5}, we consider the case when $N^{\text{BS}} < M$. We show detection results for five different NSP signals, i.e, radar waveform projected onto five different BSs. Note that, in order to achieve a detection probability of 90\%, we need 6 dB to 13 dB more gain in SNR as compared to the orthogonal waveform, depending upon which channel we select. Using Algorithms \eqref{alg:M2C_Radar} and \eqref{alg:Proj} we can select interference channel that results in minimum degradation of radar waveform and results in enhanced target detection performance with the minimum additional gain in SNR required. For example, Algorithms \eqref{alg:M2C_Radar} and \eqref{alg:Proj} would select BS\#5 because in this case NSP waveform requires least gain in SNR to achieve a detection probability of 90\% as compared to other BSs.  

In Figure \ref{fig:PdH8x2_BS_5}, we consider the case when $N^{\text{BS}} \ll M$. Similar to Figure \ref{fig:PdH4x2_BS_5} we show detection results for five different NSP signals but now MIMO radar has a larger array of antennas as compared to the previous case. In this case, in order to achieve a detection probability of 90\%, we need 3 dB to 5 dB more gain in SNR as compared to the orthogonal waveform. As in the previous case, using Algorithms \eqref{alg:M2C_Radar} and \eqref{alg:Proj} we can select interference channel that results in minimum degradation of radar waveform and results in enhanced target detection performance with the minimum additional gain in SNR required. For example, Algorithms \eqref{alg:M2C_Radar} and \eqref{alg:Proj} would select BS\#2 because in this case NSP waveform requires least gain in SNR to achieve a detection probability of 90\% as compared to the other BSs.

The above two examples demonstrates the importance of Algorithms \eqref{alg:M2C_Radar} and \eqref{alg:Proj} in selecting interference channel for radar signal projection to maximize detection probability and minimize gain in SNR required as a result of NSP of radar waveforms for spectrum sharing.

\subsection{Case 1: $\text{dim} \, \mc N [ \mbf H_i] = 2$}

\begin{figure*}
\centering
	\includegraphics[width=\linewidth]{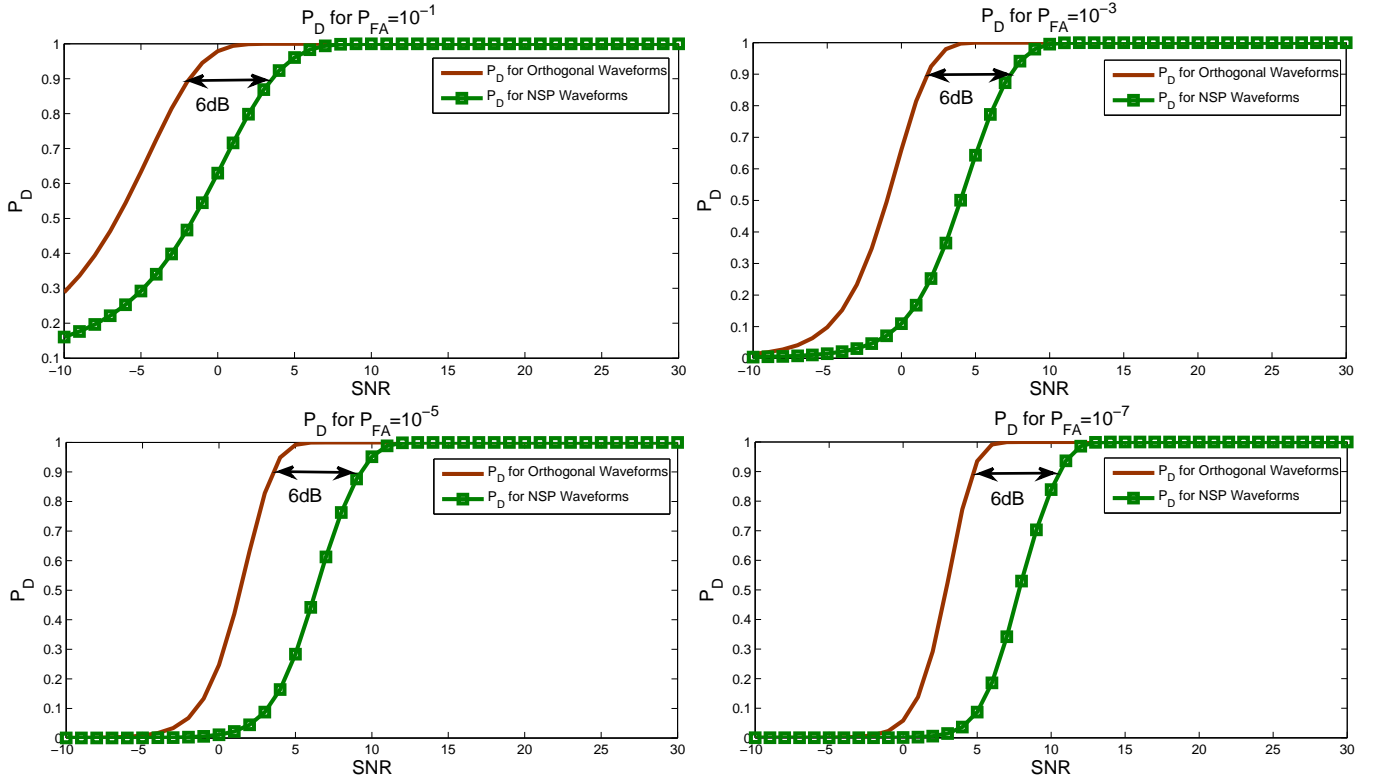} 
	\caption{'Case 1: $\text{dim} \, \mc N [ \mbf H_i] = 2$': $P_{\text{D}}$ as a function of SNR for various values of probability of false alarm $P_{\text{FA}}$, i.e., $P_{\text{FA}} = 10^{-1}, 10^{-3}, 10^{-5}$ and $10^{-7}$. The interference channel $\mbf H_i$ has dimensions $2 \times 4$, i.e., the radar has $M = 4$ antennas and the communication system has $N^{\text{BS}} = 2$ antennas, thus, we have a null space dimension of `$\text{dim} \, \mc N [ \mbf H_i] = 2$'. Note that we need 9 to 10 dB more gain in SNR for the NSP waveform to get the same result produced by the orthogonal waveform.}
		\label{fig:PdH4x2}
\end{figure*}

In Figure \ref{fig:PdH4x2}, we plot the variations of probability of detection $P_{\text{D}}$ as a function of signal-to-noise ratio (SNR) for various values of probability of false alarm $P_{\text{FA}}$. Each sub-plot represents the $P_{\text{D}}$ for a fixed $P_{\text{FA}}$. We choose to evaluate $P_{\text{D}}$ against $P_{\text{FA}}$ values of $10^{-1}, 10^{-3}, 10^{-5}$ and $10^{-7}$ when the interference channel $\mbf H_i$ has dimensions $2 \times 4$, i.e., the radar has $M = 4$ antennas and the communication system has $N^{\text{BS}} = 2$ antennas, thus, we have a null space dimension of `$\text{dim} \, \mc N [ \mbf H_i] = 2$'. When we compare the detection performance of two waveforms we note that in order to get a desired $P_D$ for a fixed $P_{\text{FA}}$ we need more SNR for NSP than orthogonal waveforms. For example, say we desire $P_D = 0.9$, then according to Figure \ref{fig:PdH4x2} we need 6  dB more gain in SNR for NSP waveform to get the same result produced by the orthogonal waveform.

\begin{figure*}
\centering
	\includegraphics[width=7in]{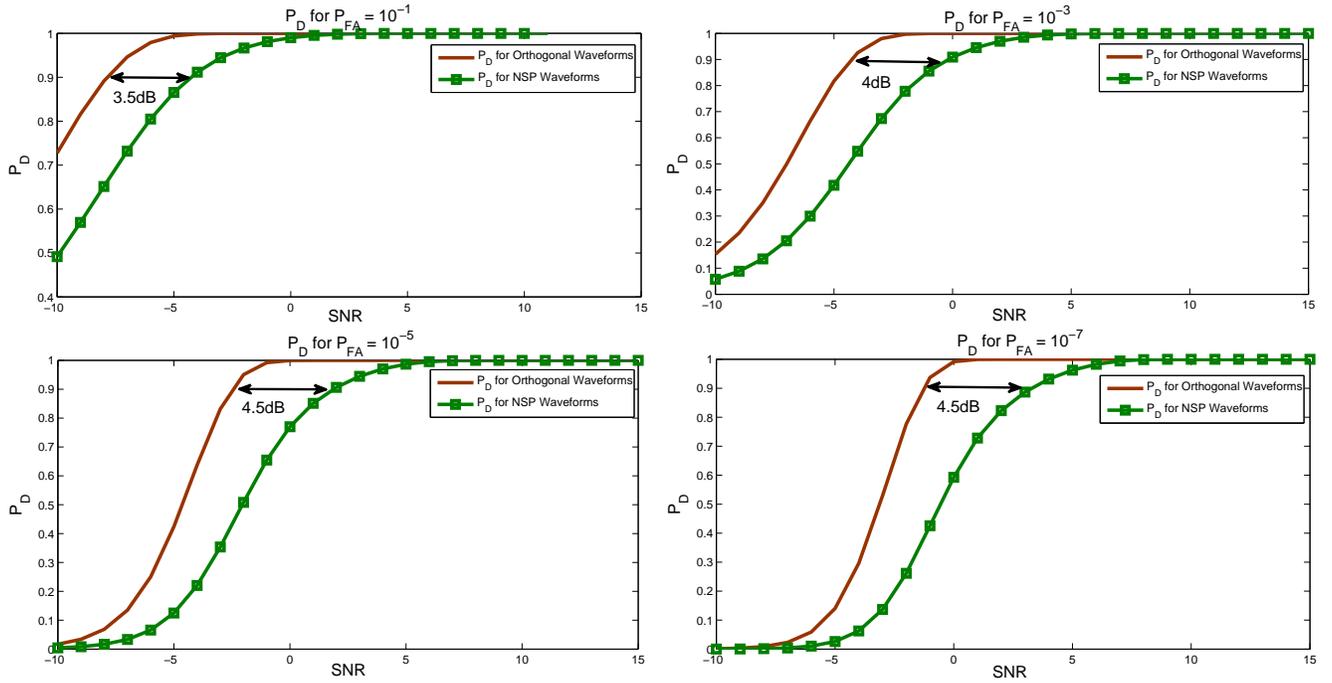} 
	\caption{'Case 2: $\text{dim} \, \mc N [ \mbf H_i] = 6$': $P_{\text{D}}$ as a function of SNR for various values of probability of false alarm $P_{\text{FA}}$, i.e., $P_{\text{FA}} = 10^{-1}, 10^{-3}, 10^{-5}$ and $10^{-7}$. The interference channel $\mbf H_i$ has dimensions $2 \times 8$, i.e., the radar has $M = 8$ antennas and the communication system has $N = 2$ antennas, thus, we have a null space dimension of `$\text{dim} \, \mc N [ \mbf H_i] = 6$'. Note that we need 3.5 to 4.5 dB more gain in SNR for the NSP waveform to get the same result produced by the orthogonal waveform.}
		\label{fig:PdH8x2}
\end{figure*}

\subsection{Case 2: $\text{dim} \, \mc N [ \mbf H_i] = 6$}

In Figure \ref{fig:PdH8x2}, similar to Figure \ref{fig:PdH4x2}, we do an analysis of $P_{\text{D}}$ against the same values of $P_{\text{FA}}$ but for interference channel $\mbf H_i$ having dimensions $2 \times 8$, i.e., now the radar has $M = 8$ antennas and the communication system has $N = 2$ antennas, thus, we have a null space dimension of `$\text{dim} \, \mc N [ \mbf H_i] = 6$'. Similar to Case 1, when we compare the detection performance of two waveforms we note that in order to get a desired $P_D$ for a fixed $P_{\text{FA}}$ we need more SNR for NSP than the orthogonal waveforms. For example, say we desire $P_D = 0.9$, then according to Figure \ref{fig:PdH8x2} we need 3.5 to 4.5 dB more gain in SNR for the NSP waveform to get the same result produced by the orthogonal waveform.


\subsection{Comparison of Case 1 and Case 2}
As expected, when SNR increases detection performance increases for both waveforms. However, when we compare the two waveforms at a fixed value of SNR, the orthogonal waveforms perform much better than the NSP waveform in detecting target. This is because our transmitted waveforms are no longer orthogonal and we lose the advantages promised by orthogonal waveforms when used in MIMO radars as discussed in Section \ref{sec:orthogonalWaveforms}, but, we ensure zero interference to the BS of interest, thus, sharing radar spectrum at an increased cost of target detection in terms of SNR.

In Case 1, in order to achieve a desired $P_D$ for a fixed $P_{\text{FA}}$ we need more SNR for NSP as compared to Case 2. This is because we are using more radar antennas, while the antennas at the BS remain fixed, in Case 2 which increases the dimension of the null space of the interference channel. This yields better detection performance even for NSP waveform. So, in order to mitigate the effect of NSP on radar performance one way is to employ a larger array at the radar transmitter.

\section{Conclusion}\label{sec:conclusion}
In the future, radar RF spectrum will be shared with wireless communication systems to meet the growing bandwidth demands and mitigate the effects of spectrum congestion for commercial wireless services. In this paper, we analyzed a similar spectrum sharing scenario between radars and cellular systems. We proposed a spectrum sharing scenario in which a MIMO radar is sharing spectrum with a cellular system. We proposed algorithms for interference channel selection and projection of radar waveform onto the selected interference channel in order to mitigate interference to the selected BS. We evaluated the detection performance of spectrum sharing MIMO radars. We formulated the statistical detection problem for target detection and used generalized likelihood ratio test to decide about the presence of target when using orthogonal waveforms and null-space projected (NSP) waveforms. We showed that by using our spectrum sharing and projection algorithms the radar can maximize target detection probability and 
minimize additional gain in SNR required to detect the target. We showed that when using the NSP waveforms, the detection performance degrades as compared to the orthogonal waveforms and we need more SNR to detect reliably. Our results showed that, about 6 dB of gain in SNR is required when $ N^{\text{BS}} < M $ and 3.5 to 4.5 dB of gain in SNR is required when $ N^{\text{BS}} \ll M$ when NSP waveforms are used instead of orthogonal waveforms for spectrum sharing. Our analysis showed that this degradation in performance can be mitigated by using a larger array at the MIMO radar transmitter.

\bibliographystyle{ieeetr}
\bibliography{IEEEabrv,ProbOfDetection}
\end{document}